\documentclass[12pt]{iopart}

\usepackage{latexsym,epsfig, amsfonts}

\expandafter\let\csname equation*\endcsname\relax
\expandafter\let\csname endequation*\endcsname\relax
\usepackage{amsmath}
\usepackage{xcolor}
\usepackage{bm}

\def\kgr{{\bf p} }

\def\0gr{{\bf 0} }

\def\beq{\begin{equation}}
\def\eeq{\end{equation}}

\begin{document}

\title{Casimir forces for the ideal Bose gas in anisotropic optical lattices: the effect of alternating sign upon varying dimensionality}

\author{Marek Napi\'{o}rkowski}
\address{Institute of Theoretical Physics, Faculty of Physics, University of Warsaw, Pasteura 5, 02-093 Warsaw, Poland}
\ead{marek.napiorkowski@fuw.edu.pl}

\author{Pawel Jakubczyk}
\address{Institute of Theoretical Physics, Faculty of Physics, University of Warsaw, Pasteura 5, 02-093 Warsaw, Poland}
\ead{pawel.jakubczyk@fuw.edu.pl}


\begin{abstract}
We analyze the thermodynamic Casimir effect occurring in a gas of non-interacting bosons confined by two parallel walls with a strongly anisotropic dispersion inherited from an underlying lattice. In the direction perpendicular to the confining walls the standard quadratic dispersion is replaced by the term $|{\bf p}|^{\alpha}$ with $\alpha \geq 2$ treated as a parameter. We derive a closed, analytical expression for the Casimir force depending on the dimensionality $d$ and the exponent $\alpha$, and analyze it for thermodynamic states in which the Bose-Einstein condensate is present. For $\alpha\in\{4,6,8,\dots\}$ the exponent governing the decay of the Casimir force with increasing distance between the walls becomes modified and  the Casimir amplitude $\Delta_{\alpha}(d)$ exhibits oscillations of sign as a function of $d$. Otherwise we find that $\Delta_{\alpha}(d)$ features singularities when viewed as a function of $d$ and $\alpha$. Recovering the known previous results for the isotropic limit $\alpha=2$ turns out to occur via a  cancellation of singular terms. 
\end{abstract}

\maketitle

\section{Introduction} 
The Casimir effect has been receiving considerable interest over the last years in a range of versatile physical contexts approached from theoretical, numerical and experimental points of view \cite{Krech_book,Kardar_1999, Brankov_book, HHGDB2008, Klimchitskaya_2009, Gambassi_2009, Maciolek_2018}. In addition to proving its historical relevance as one of the key testable QED predictions \cite{Lamoreaux_1997}, it appeared to be of strikingly broad universality and is nowadays invoked in a diversity of situations ranging from biological cells \cite{Machta_2011, Machta_2012} to theories of fundamental interactions \cite{Masuda_2009, Klimchitskaya_2012} and cosmology \cite{Mello_2012}. In recent years its existence and theoretically predicted properties have also been firmly established on both qualitative and quantitative levels by means of high-precision experiments \cite{Garcia_1999, Garcia_2002, HHGDB2008, Klimchitskaya_2009, Maciolek_2018}  as well as extensive numerical simulations (see e.g. \cite{Vasilyev_2007, Hasenbush_2012, Hasenbush_2015}). 

Among the most obvious characteristics of the Casimir force is its sign. It is well recognized and supported by exact results that the fluctuation-induced Casimir force acting between (non-magnetic) bodies related by a reflection (i.e. being mirror images of each other) immersed in a homogeneous medium is bound to be attractive \cite{Li_1997, Kenneth_2006} irrespective of the nature of these bodies (which in particular pertains the boundary conditions), or properties of the uniform medium. A number of studies \cite{Soyka_2008, Nellen_2009, Jakubczyk_2016_2, Flachi_2017, Faruk_2018, Jiang_2019} proposed ways around the assumptions underlying the exact statements, the most direct ones invoking different boundary conditions at each of the two macroscopic objects between which the force acts, or non-uniformities of the fluctuating medium. Note also, that the Casimir force (including the sign) may in addition exhibit different properties depending on the statistical ensemble being used \cite{Gross_2016}. 

It was however recently observed \cite{Burgsmuller_2010}, that the generically attractive nature of the Casimir interactions is not preserved for anisotropic systems characterized by dispersion relations deviating from a quadratic form. Indeed, the exact proofs provided in Refs.~\cite{Li_1997, Kenneth_2006} employ a particular form of the field propagator and do not apply to this case. As was for the first time indicated in Ref.~\cite{Burgsmuller_2010} for the general context of Lifshitz transitions \cite{Hornreich_1975, Chaikin_book, Diehl_2002, Essafi_2012}, and somewhat later in Ref.~\cite{Lebek_2020} for imperfect Bose gases on anisotropic lattices, the Casimir force may be repulsive for situations where the dispersion relation is quartic in some of the directions. A further study \cite{Lebek_2021} exhibited the key role of spatial dimensionality $d$ and demonstrated that the sign of the Casimir interaction alternates upon varying $d$. In particular, for the so-called uniaxial case characterized by a quartic dispersion in a single direction, it is strictly zero at $d$ corresponding to even natural values  ($d\in\{4,6,8,\dots\}$). This deviates very far from the standard situations, where dimensionality has literally no impact on the sign of the Casimir force \cite{Napiorkowski_2013}. The imperfect (mean-field) Bose gas addressed in Refs.~\cite{Lebek_2020, Lebek_2021} is a representative  of the anisotropic vectorial $N\to\infty$ universality class and these unusual features are  expected to occur in a much broader class of systems, including some magnets \cite{Dohm_2021}, liquid crystals \cite{Singh_2000} and non-uniform superfluids of the Fulde-Ferrell-Larkin-Ovchinnikov class \cite{fulde_superconductivity_1964, larkin_nonuniform_1965, radzihovsky_fluctuations_2011}. The findings reported in Ref.~\cite{Burgsmuller_2010} and Refs. \cite{Lebek_2020, Lebek_2021} call for further studies concerning examination of the generality of the predicted picture as well as its further resolution. 

In the present paper we readdress this problem by analyzing the non-interacting Bose gas exhibiting a non-isotropic dispersion inherited from an underlying lattice. We find effects very similar to those reported in Ref.~\cite{Lebek_2020} for the imperfect (mean-field) Bose gas pointing towards robustness of the general picture. Taking advantage of the extreme simplicity of the studied model, we additionally develop a theoretical setup allowing for a continuous deformation of the dispersion making it possible to continuously  interpolate between the cases of quadratic 
 and quartic 
 dispersions $\epsilon ({\bf p})$ corresponding to Casimir forces of different sign. Namely, we consider 
 \beq 
 \epsilon ({\bf p})\sim |{\bf p}|^\alpha
 \eeq 
  for ${\bf p}$ oriented in the direction perpendicular to the macroscopic objects between which the Casimir force acts, with the parameter $\alpha\geq 2$ arbitrary. The applied procedure borrows the spirit of the successful techniques such as the $\epsilon$-expansion or the $1/N$-expansion, where one considers the natural quantities such as the spatial dimensionality $d$ or the number of spin components $N$ as a continuous variable. Surprisingly, we find that $\alpha$ cannot be varied continuously between $2$ and $4$ without encountering singularities of the Casimir energy. 
  Conversely, at $\alpha$ fixed the Casimir amplitude turns out to be a non-singular function of $d$ only for $\alpha \in \{2,4,6,\dots\}$.   
  
  Note in addition, that interesting phenomena (not discussed in this paper) occur also in situations where $\alpha=2$ but the coefficients governing the amplitude of the dispersion relation depend on the momentum direction \cite{Dohm_2010}.  

The outline of the paper is as follows: In Sec.~2 we introduce the model and in Sec.~3 we summarize its relevant bulk properties. Sec.~4 contains our key results for the Casimir force disclosing an affinity between the ideal and imperfect Bose gases concerning this aspect. We evaluate a closed, analytical form of the Casimir amplitude $\Delta_\alpha (d)$ at arbitrary dimensionality $d$ and the parameter $\alpha$ and scan its dependence on $d$ for the physically most relevant cases of $\alpha\in\{2, 4, 6,\dots\}$, revealing oscillations of the sign of $\Delta_\alpha (d)$ upon varying $d$ for $\alpha\in\{4,6,\dots\}$. We subsequently analyze continuous $\alpha$ and demonstrate the impossibility of connecting the standard situation with $\alpha=2$ with that corresponding to $\alpha=4$ via a continuous deformation of the dispersion (varying $\alpha$). In particular we show that at fixed dimensionality $d=3$ an interpolation between $\Delta_{2 }(3)<0$ and $\Delta_{4} (3)>0 $ by varying $\alpha$ encounters a singularity of $\Delta_\alpha (3)$ at $\alpha=3$.
In Sec.~5 we summarize our results.    

\section{Ideal Bose gas on a lattice} 
Consider a prototype system of non-interacting, spinless bosons on a $d$-dimensional hypercubic lattice, governed by the Hamiltonian 
\beq 
\label{Hamiltonian}
\hat{H} = -\sum_{({\bf x}, {\bf y})}t_{{\bf x} {\bf y}}a^\dagger_{\bf x} a_{\bf y} =  \sum_{\bf p}\epsilon ({\bf p})\hat{n}_{\bf p}\;,
\eeq 
where ${\bf x}$ and ${\bf y}$ label the points of the lattice, $({\bf x}, {\bf y})$ stands for a pair of lattice points,    
$\hat{n}_{\bf p}=a_{\bf p}^\dagger a_{\bf p}$ denotes  the occupation number operator of a single-particle state labeled by momentum $\bf p$.  
 A general tight-binding dispersion is given as 
 \beq   
 \epsilon ({\bf p})=\sum_{\bf x} 2t_{\bf x}[1-\cos ({\bf p x}/\hbar)]  
 \eeq
with ${\bf x}$ running over the lattice sites and $t_{\bf x}=t_{\bf 0 x}$ denoting the corresponding hopping amplitudes. Only the asymptotic behavior of $\epsilon ({\bf p})$ at $|\bf p|$ small (which is usually taken to be quadratic) is relevant for the universal critical properties at Bose-Einstein condensation. As was discussed in Ref.~\cite{Jakubczyk_2018} it is however possible to tune the  hopping amplitudes  $\{t_{\bf x}\}$ so that the coefficient of the leading quadratic term in the  small $|\bf p|$ expansion of $\epsilon ({\bf p})$ vanishes, which promotes the subleading terms (typically quartic in $\bf p$) to the dominant contributions and makes the dispersion anomalously flat. Such a tuning procedure can be performed independently in each of the $d$ spatial directions. It may also be pushed further leading to a cancellation of an arbitrary number of leading terms in the small-$|{\bf p}|$ expansion of $\epsilon ({\bf p})$ in each of the directions independently, such that the leading contribution is governed by an even, arbitrarily high power of momentum.
 In an experimental situation involving a gas of ultracold bosons on an optical lattice,  this can be achieved (at least in principle) by manipulating the optical lattice parameters via  Feshbach resonances. An anisotropic tunability of such lattices was indeed considered in some recent experimental situations \cite{Greif_2013, Imriska_2014}.   Here we are primarily interested in the setup, where the dispersion is quadratic along $(d-1)$ directions parallel to the walls and varies as $|p_d|^\alpha$ in the 'special' $d$-th direction perpendicular to the walls. A generalization to the case of more than one such special directions is also possible. We note that, at least for the presently considered case of a hypercubic lattice, only even natural values of the parameter $\alpha$ are admissible. We shall however treat this quantity as an arbitrary  variable and investigate in particular if a continuous interpolation between $\alpha =2$ and $\alpha =4$ is possible. For the sake of  simplicity,  we shall expand the dispersion around ${\bf p}=0$ as follows: 
\begin{equation} 
\label{asssym}
\epsilon({\bf p}) \to \tilde{\epsilon}({\bf p})  = \sum_{i=1}^{d-1} t_0 p_i ^2 +  t|p_d| ^\alpha\;, 
\end{equation} 
replacing the expression for $\epsilon (\bf{p})$ with its low-momentum asymptotic form $\tilde{\epsilon}({\bf{p}})$ and assuming the hopping amplitudes  have been chosen so that the dispersion is quadratic in $(d-1)$ directions and varies as $|p_d|^\alpha$ in the remaining $d$-th direction. We also assume  $t_0>0$, $t>0$. 
The volume of the $d$-dimensional hypercube is  $V=L^{d-1}D$, where $L\gg D\gg l_{mic}$ and $l_{mic}$ encompasses all the microscopic length scales present in the system. The quantity $D$ measures the system extension in the special $d$-th direction [see Eq.~(\ref{asssym})]. We impose Neumann boundary conditions in all directions which leads to the following momentum spectrum  
\begin{align} 
p_{i} = \frac{\hbar \pi}{L} \, n_{i}\;, \hspace{1cm} n_{i} &= 0,1, 2,.... \;, \hspace{1cm} i=1,...,d-1\;; \nonumber \\  
p_{d} =  \frac{\hbar \pi}{D} \, n_{d}\;,  \hspace{1cm} n_{d} &= 0,1, 2,....  \; .
\end{align}
This choice is complementary to the one discussed in Refs.~\cite{Lebek_2020, Lebek_2021} for the mean-field Bose gas, where periodic boundary conditions were used. As compared to the periodic b.c. the Neumann boundary conditions lead to the appearance of a non-zero interfacial tension term and to the modifications of some numerical coefficients, but, as we show below, do not alter the qualitative properties of the Casimir energy in any respect. Note that the more realistic choice of Dirichlet boundary conditions leads to subtleties when considering the approach to the phase hosting the Bose-Einstein condensate \cite{Napiorkowski_2020}, which we do not discuss here. 

In order to analyze the Casimir interactions we employ the grand canonical framework. The corresponding free energy $\Omega(T,\mu,L,D)$ is given as
\beq
\label{gcfe1} 
\Omega(T,\mu,L,D) = k_{B} T \sum\limits_{p_{1}} ... \sum\limits_{p_d}\
\log \left(1 - e^{\beta[\mu - \epsilon ({\bf p})]
}\right)\;,
\eeq 
where $T=\beta^{-1}/k_B$ denotes the temperature and $\mu$ is the chemical potential.  The relevant quantity for evaluating the Casimir forces is the surface free energy per unit area obtained by subtracting from $\Omega(T,\mu,L,D)$ the bulk term $ - L^{d-1}D\, p(T,\mu)$, where $p(T,\mu)$ denotes the ideal Bose gas pressure. \\
\section{Bulk properties}
Below we summarize the relevant bulk properties of the system defined in Sec.~2. The bulk grand canonical free energy density $\omega_{b}(T,\mu)$ is evaluated in the thermodynamic limit 
\beq
\omega_{b}(T,\mu) = \lim\limits_{L \rightarrow \infty} \frac{\Omega(T,\mu,L,L)}{L^d}
\eeq
and takes the following form
\beq
\label{bfed1}
\omega_{b}(T,\mu) = - \frac{k_B T}{\lambda_{1}^{d-1} \lambda_{2}} \,\int\limits_{0}^{\infty} dx \,g_{\frac{d+1}{2}}\left(e^{\beta\mu - x^\alpha}\right) = - \frac{k_B T}{\lambda_{1}^{d-1} \lambda_{2}} \Gamma (1+\alpha^{-1}) g_{\frac{\alpha(d+1)+2}{2\alpha}}\left(e^{\beta\mu}\right)  \;,
\eeq
where 
\beq 
\label{Bose_f}
g_{\kappa}(z) = \sum\limits_{n=1}^{\infty} \frac{z^n}{n^{\kappa}}
\eeq 
 is the Bose function \cite{Ziff_1977}. The quantities 
\beq 
\lambda_{1} = \left(\frac{h^2 \beta t_{0}}{\pi}\right)^{1/2} \quad  \textrm{and} \quad  \lambda_{2} =  \frac{h\left(\beta t\right)^{1/\alpha}}{2} 
\eeq
represent the analogues of the thermal de Broglie wavelength $\lambda = \frac{h}{\sqrt{2\pi mk_{B}T}}$; in particular $\lambda_{1}=\lambda$ for $t_{0}=1/(2m)$. Note that the standard, continuum noninteracting Bose gas is recovered by taking $\alpha=2$, $t_{0}=t=1/(2m)$. In this case $\lambda_{2} = \frac{\pi^{1/2}}{2} \lambda$. When evaluating the momentum integrals 
leading to Eq.~(\ref{bfed1}) we employed the model dispersion relation $\tilde{\epsilon}({\bf p})$, see Eq.~(\ref{asssym}).  The approximation leading to Eq.~(\ref{asssym}) does not influence our conclusions concerning the universal properties in the vicinity of the transition (and also in the low-$T$ phase)  since the dominant contributions responsible for the critical singularities are controlled by low momenta \cite{Jakubczyk_2018}. We will consequently remain at this approximation level in what follows.

The analysis of Bose-Einstein condensation in the considered anisotropic system follows along the standard lines \cite{Ziff_1977}. The Bose-Einstein condensate is present at $\mu=0$ and $T < T_{c}$. Its existence is proved by considering the thermodynamic limit $V = L^{d} \rightarrow \infty$ simultaneously with the limit $\mu \rightarrow 0^-$ such that 
\beq
\mu = \epsilon_{G}^{N} -  k_{B}T \log \left(1 - \frac{1}{V \rho_{0}}\right) \;. 
\eeq 
In the above formula $\rho_{0}$ denotes the condensate density while $\epsilon_{G}^{N} $ is the single particle ground state energy evaluated   for Neumann boundary conditions, $\epsilon_{G}^{N} = 0$. In particular, the expression for the particle number density 
\beq
\label{dens01}
\rho = \frac{\langle N\rangle}{V} = \frac{1}{V} \frac{1}{e^{-\beta\mu}-1} + \frac{\Gamma (1+\alpha^{-1})}{\lambda_{1}^{d-1} \lambda_{2}} \, g_{\frac{\alpha (d-1)+2}{2\alpha}}(e^{\beta\mu})
\eeq 
takes in the above limit the standard form
\beq
\label{dens02}
\rho = \rho_{0} + \rho_{c}(T)\;,
\eeq
where $\rho_{0}$ denotes the condensate density and $\rho_{c}(T)$ is the critical density. The average particle number $\langle N \rangle$ is related to $\omega_b$ via $V \langle N \rangle=-\partial \omega_b/\partial\mu$.  Eq.~(\ref{dens02}) allows for the evaluation of the condensate density 
$\rho_{0}(T,\rho)$. The critical density $\rho_{c}(T)$ is defined for $d > 3-\frac{2}{\alpha}$ and is given by 
$\rho_{c} = \frac{\Gamma(1+\alpha^{-1})\zeta\left(\frac{\alpha (d-1)+2}{2\alpha}\right)}{\lambda_{1}^{d-1} \lambda_{2}}$, where $\zeta(\kappa)=g_\kappa (1)$ denotes the Riemann zeta function. Alternatively, the critical temperature $T_{c}$ as a function of density $\rho$ may be obtained from 
\beq 
 \rho = \frac{\Gamma (1+\alpha^{-1}) \zeta\left(\frac{\alpha (d-1)+2}{2\alpha}\right)}{\lambda_{1,c}^{d-1} \lambda_{2,c}} 
 \eeq 
  and thus $T_{c}(\rho) \sim \rho^{\psi}$, 
where $\psi = \frac{2\alpha}{2+\alpha (d-1)}$ \cite{Lebek_2020}. Upon putting $\alpha=2$ or $\alpha=4$ this exponent coincides with the corresponding values obtained for the imperfect Bose gas \cite{Lebek_2020, Jakubczyk_2013_2}.

  Note that the above expressions follow from the dispersion relation $\tilde{\epsilon}({\bf p})$ in Eq.~(\ref{asssym}) which we assume to define our model system. Thus it serves  as the basis of our analysis for arbitrary temperatures and densities. If, however, the dispersion relation $\epsilon({\bf p})$ had been implemented instead of its asymptotic form $\tilde{\epsilon}({\bf p})$, the above formulae would remain correct only in the low-$T$ limit and the universal power law governing the behavior of $T_c$ would hold only asymptotically for small $T_c$ \cite{Jakubczyk_2018}.
\section{Casimir forces}
We now consider the Casimir force, which is evaluated according to the definition
\beq
\label{Cas001}
{F}_{C}(T,\mu,D) =  -\frac{\partial \omega_{s}(T,\mu,D)}{\partial D} \;,
\eeq
where 
$\omega_{s}(T,\mu,D)$ denotes the surface free energy density
\beq
\label{pes1}
\omega_{s}(T,\mu,D) = \lim\limits_{L \rightarrow \infty} \frac{\Omega(T,\mu,L,D) - L^{d-1}D \,\omega_{b}(T,\mu)}{L^{d-1}} \quad.
\eeq 
It follows from equations (\ref{pes1}), (\ref{gcfe1}), and (\ref{bfed1}) that 
\begin{align}
\label{pes2}
\omega_{s}(T,\mu,D) \, \frac{\lambda_{1}^{d-1}}{k_{B}T} &= - \sum\limits_{n=0}^{\infty} \,g_{\frac{d+1}{2}} 
\left(e^{\beta\mu - \left(\frac{\lambda_{2} n}{D}\right)^\alpha} \right) \, + \, \frac{D}{\lambda_{2}} \, \int\limits_{0}^{\infty} dx \,g_{\frac{d+1}{2}}\left(e^{\beta\mu - x^\alpha}\right) = \nonumber \\
&= 2\, \sigma^{N} \, \frac{\lambda_{1}^{d-1}}{k_{B}T}  - 2\sum\limits_{p=1}^{\infty} \int\limits_{0}^{\infty} dx \,g_{\frac{d+1}{2}}\left(e^{\beta\mu - \left(\frac{\lambda_{2} x}{D}\right)^\alpha}\right)\, \cos(2 \pi p x) \;. 
\end{align}
In evaluating the above expression we used the non-expanded form of the Euler-Maclaurin formula  \cite{Napiorkowski_2014} 
\beq
\label{summ}
\sum\limits_{n=1}^{\infty} \varphi(n) = \int\limits_{0}^{\infty} dx \varphi(x) - \frac{1}{2} \varphi(0) + 2 \sum\limits_{p=1}^{\infty} \int\limits_{0}^{\infty} dx \varphi(x) \cos(2\pi p x) \;.
\eeq
The first term in the second line of Eq.~(\ref{pes2}) represents the $D$-independent contribution to the surface free 
energy density given by the surface tension coefficients 
\beq
\label{st1}
\sigma^{N}(T,\mu) =  - \frac{k_{B}T}{4 \lambda_{1}^{d-1}} \,g_{\frac{d+1}{2}}\left(e^{\beta\mu}\right) 
\eeq
which is negative (the superscript $N$ stands for Neumann b.c.). An identical expression for the surface tension coefficient was previously derived  for the "standard"  (i.e. corresponding to $\alpha=2$) situation
 (with Neumann boundary conditions) \cite{Napiorkowski_2014, Martin_2006}. This indicates that the structure of the surface tension coefficient is insensitive to the fact that the contribution from the $d$-direction to the dispersion relation in Eq.~(\ref{Hamiltonian}) is of the form $p_{d}^\alpha$ rather then the "standard" case $p_{d}^2$.  We also recall that in the case of "standard" ideal Bose gas with Dirichlet boundary conditions one observes a positive surface tension coefficient  
$\sigma^{D} = - \sigma^{N} >0$  \cite{Napiorkowski_2020, Napiorkowski_2014, Martin_2006}, while for the periodic boundary conditions $\sigma^{P}=0 $. \\

We now concentrate on the $D$-dependent contribution to the surface free energy density 
\beq 
\delta\omega_{s}(T,\mu,D) = \omega_{s}(T,\mu,D) -2 \sigma^N 
\eeq
 evaluated for particular thermodynamic states defined by $\mu=0$ and $T < T_{c}$, i.e., corresponding to the phase in which the condensate is present. This is given by 
\beq 
\delta\omega_{s}(T,0,D) = \frac{k_{B}T}{\lambda_{1}^{d-1}} \, \Phi_d^{(\alpha)}(T,D) \;,
\eeq  
where  
\beq
\label{Fi1}
\Phi_d^{(\alpha)}(T,D) = - 2\sum\limits_{p=1}^{\infty} \int\limits_{0}^{\infty} dx \,g_{\frac{d+1}{2}}\left(e^{- \left(\frac{\lambda_{2} x}{D}\right)^\alpha}\right)\, 
\cos(2 \pi p x) \;. 
\eeq
The above function encodes the entire structure of the Casimir energy $\delta\omega_{s}(T,0,D)$ at arbitrary $d>3-\frac{2}{\alpha}$ and $\alpha \geq 2$.  For the standard isotropic case with $\alpha=2$ this function displays, for large distances $D$, a power-law decay 
$\delta\omega_{s}(T,0,D)/k_{B}T \,= \, \Delta_{2}(d)/D^{d-1}$ with a universal amplitude $\Delta_{2}(d)$ \cite{Napiorkowski_2014, Martin_2006}. A detailed inspection presented below indicates however a complex and interesting behavior  of $\Phi_d^{(\alpha)} (T,D)$ for $\alpha\neq 2$. In particular recall that a related study performed in Ref.~\cite{Lebek_2021} for the mean-field (imperfect) Bose gas with fixed $\alpha = 4$ revealed a change of sign of the Casimir energy (and the Casimir force as well) as function of the dimensionality, such that the force is strictly zero at $d\in \{4,6,8,\dots\}$, repulsive for $d\in (\frac{5}{2},4)\cup (6,8)\cup \dots$ and attractive for $d\in (4,6)\cup (8,10)\cup \dots$. This is completely different as compared to the standard situation ($\alpha =2$), where the Casimir force is attractive for any $d$.  An essential objective of the current analysis is to check whether similar oscillations of the Casimir amplitude upon varying dimensionality at $\alpha=4$ occur also for the the presently analyzed case of the ideal Bose gas. Moreover, the validity of the expression derived above is not restricted to $\alpha\in\{2,4\}$ and one may consider the higher physically admissible values of $\alpha$ ($\alpha\in\{6,8,10,\dots\}$) as well as  evolve the system by varying the parameter $\alpha$ continuously at arbitrary dimensionality $d$. 

With these goals in mind we define 
\beq 
\label{Psi_def}
\Psi_\alpha (z) =\int_0^\infty dx e^{-x^\alpha} \cos(zx) 
\eeq
and use the series representation of the Bose function $g_{\frac{d+1}{2}}(z)$ [Eq.~(\ref{Bose_f})] and the Euler-Maclaurin formula [Eq.~(\ref{summ})] 
to transform Eq.~(\ref{Fi1}) to the following form: 
\begin{align}
\Phi_d^{(\alpha)}(T,D) &= \Bigg[-\frac{2\alpha}{(2\pi)^{1+\frac{\alpha(d-1)}{2}}}\zeta\left(1+\frac{\alpha(d-1)}{2}\right)\int_0^\infty dz z^{\frac{\alpha(d-1)}{2}}\Psi_\alpha(z) + \nonumber \\
&-4\sum_{p=1}^{\infty}\sum_{m=1}^{\infty}\int_0^\infty dz \Psi_\alpha (2\pi p z^{-1/\alpha}) z^{-\frac{2+\alpha (d+1)}{2\alpha}}\cos (2\pi m z s^\alpha)\Bigg] s^{-\alpha(d-1)/2}\;,
\end{align}
where we introduced the dimensionless distance $s=D/\lambda_2$. In the limit of large distances ($s\gg 1$) the first term dominates the right-hand side of the above equation and we obtain 
\beq 
\label{Cas_en}
\frac{\delta\omega_s(T,0,D)}{k_B T} = \frac{\Delta_\alpha (d)}{D^{\alpha(d-1)/2}}\;, 
\eeq
where 
\beq 
\label{Cas_amp}
\Delta_\alpha (d) = -\left(\frac{\lambda_2^{\alpha/2}}{\lambda_1}\right)^{d-1} \frac{2\alpha\zeta (1+\alpha (d-1)/2)}{(2\pi)^{1+\alpha (d-1)/2}}\int_0^\infty dz z^{\alpha (d-1)/2} \Psi_\alpha (z)\;.
\eeq
As is transparent from Eq.~(\ref{Cas_en}), the value of the exponent governing the Casimir energy decay as function of $D$ can be altered by manipulating the parameter $\alpha$. For $\alpha=2$ we recover the standard behavior given as $\delta \omega_s\sim 1/D^{d-1}$. For $\alpha =4$ we obtain $\delta\omega_s\sim 1/D^{2d-2}$, which coincides with the power law predicted for the imperfect Bose gas in Refs.~\cite{Lebek_2020, Lebek_2021} as well as for the Lifshitz points in Ref.~\cite{Burgsmuller_2010}. Note that the modification of the exponent governing the power-law decay is accompanied by an appearance of a non-universal, dimensionful factor controlled by the lengthscales $\lambda_1$ and $\lambda_2$ (which depend on the hopping parameters $t_0$ and $t$). Also observe that a non-universal  factor remains present for anisotropic situations with $\alpha=2$. 
This is in line with the previous studies (in particular Ref.~\cite{Burgsmuller_2010}), which emphasized that the {\it scaling functions} for the Casimir energy in anisotropic systems  retain their universal character only after a non-universal dimensionful quantity is factored out. As indicated by Eq.~(\ref{Cas_en}), the exponent governing the decay of $\delta\omega_s$ linearly interpolates between the two previously analyzed values $\alpha=2$ and $\alpha=4$. Note however that Eq.~(\ref{Cas_en}) holds also for higher values of $\alpha$. The dependence of the decay exponent on $\alpha$ was, to our knowledge, not addressed in any of the previous studies. The behavior described by Eq.~(\ref{Cas_en}) may seem somewhat counterintuitive, since higher values of $\alpha$ obviously imply softer fluctuations, but turn out to yield a faster decay of the Casimir interactions. Note however that, as we show below, the corresponding Casimir amplitude typically rapidly grows upon elevating $\alpha$ at fixed $d$.

The information concerning the sign of the Casimir interaction is contained in the integral 
\beq 
\label{I_int_1}
I_\alpha (d) = \int_0^\infty dz z^{\alpha (d-1)/2} \Psi_\alpha (z)=\int_0^\infty dz z^{\alpha (d-1)/2}\int_0^\infty dx e^{-x^\alpha}\cos (zx)\;,
\eeq
occurring in Eq.~(\ref{Cas_amp}), which we now analyze in detail. A change of the order of integration supplemented by a change of integration variables leads to the following factorized form of this integral: 
\beq 
\label{I_int_2}
I_\alpha (d) = \frac{1}{\alpha} \, \int_0^\infty dw e^{-w}w^{-(d+1)/2}\int_0^{\infty} dz z^{u-1}\cos z\;, 
\eeq
where we introduced $u=1+\frac{\alpha (d-1)}{2}$. Note that the above change of integration order is not legitimate for arbitrary values of $d$, since the resulting integral of Eq.~(\ref{I_int_2}) converges only for $u\in (0,1)$. Our way to proceed is to evaluate the integral in question for $u\in (0,1)$ and use the obtained result as the basis of analytic continuation into the physically interesting domain. From Eq.~(\ref{I_int_2}) we straightforwardly obtain 
\beq 
\label{Ia1}
I_\alpha (d) = \frac{1}{\alpha} \, \Gamma \left(\frac{1-d}{2}\right) \Gamma (u) \cos\left(\frac{\pi u}{2}\right)\;.
\eeq
We note that the function in Eq.~(\ref{Ia1}) reproduces results obtained by direct numerical evaluation  of the double integral in Eq.~(\ref{I_int_1}). 
Inserting the above expression for $I_{\alpha}(d)$ into Eq.~(\ref{Cas_en}) leads to 
\beq 
\label{Cas_amp_fin}
\Delta_\alpha (d) = 2\left(\frac{\lambda_2^{\alpha/2}}{\lambda_1}\right)^{d-1} \frac{\zeta (1+\alpha (d-1)/2)}{(2\pi)^{1+\alpha (d-1)/2}} \Gamma\left(1+\frac{\alpha (d-1)}{2}\right)\Gamma\left(\frac{1-d}{2}\right)\sin\left(\frac{\pi\alpha(d-1)}{4}\right)\;,
\eeq
which is the major result of the present paper. It constitutes our final expression for the Casimir amplitude, applicable for arbitrary $\alpha\geq 2$ and the dimensionality $d>3-\frac{2}\alpha$ within the phase hosting the Bose-Einstein condensate, i.e. for $\mu=0$ and $T<T_c$. It  sheds light on the origin of the oscillations previously reported for the mean-field Bose gas for $\alpha=4$  and reveals a subtle and singular nature of the quantity $\Delta_\alpha (d)$ treated as a function of $d$ and $\alpha$. We also note that the integral given by Eq.~(\ref{I_int_1}), specified to $\alpha=4$ appeared in the analysis of the imperfect Bose gas in Ref.~\cite{Lebek_2021}. Here we succeeded in expressing it via the Euler gamma and sinus functions in the more general situation involving arbitrary $\alpha$. This also demonstrates that the two Casimir amplitudes are in fact the same up to a factor depending on the boundary conditions but not on $d$ and $\alpha$.   

By inspecting Eq.~(\ref{Cas_amp_fin}) we first observe the presence of the oscillatory factor $\sin\left(\frac{\pi\alpha(d-1)}{4}\right)$, which vanishes for $\alpha (d-1) = 4n$ ($n\in \mathbb{Z}$) and the factor $\Gamma\left(\frac{1-d}{2}\right)$ which is singular for odd integer values of  $d$ (i.e for $d\in \{3,5,7,\dots\}$). Both $\sin\left(\frac{\pi\alpha(d-1)}{4}\right)$ and $\Gamma\left(\frac{1-d}{2}\right)$ change sign when varying $d$. In the former case this happens by crossing zero, in the latter, by crossing a singularity. The interplay between these two yields interesting behavior, which very strongly depends on the value of $\alpha$. 

We now make a number of observations concerning the properties of the expression for $\Delta_\alpha (d)$ at different (fixed) values of $\alpha$. \\

  1) For $\alpha=2$ (and exclusively for this case!) the zeroes of $\sin\left(\frac{\pi\alpha(d-1)}{4}\right)$ precisely correspond to the singular points of $\Gamma\left(\frac{1-d}{2}\right)$ such that the limiting values of their product is always finite and negative. In consequence: the Casimir amplitude $\Delta_{\alpha=2} (d)$ has a fixed (negative) sign irrespective of $d$, which is in line with the well-known previous results \cite{Martin_2006, Napiorkowski_2014}. In particular, by restricting to the isotropic continuum case we recover for $d=3$ the well-known result $\Delta_2 (3)=-\zeta (3)/(8\pi)$ \cite{Martin_2006}. \\
  
 2) For $\alpha\in \{4, 6,8,\dots\}$ each singular point of $\Gamma\left(\frac{1-d}{2}\right)$ falls at a zero of $\sin\left(\frac{\pi\alpha(d-1)}{4}\right)$ and their product gives a finite, nonzero limiting value. The converse is however not true: there are zeroes of $\sin\left(\frac{\pi\alpha(d-1)}{4}\right)$ which do not correspond to singular points of $\Gamma\left(\frac{1-d}{2}\right)$ and thus the amplitude $\Delta_{\alpha} (d)$ vanishes at such special points. In fact, each interval $d\in (2n-1,2n+1)$ with $n\in{2,3,4,\dots}$ spanned between two singular points of $\Gamma\left(\frac{1-d}{2}\right)$ hosts the number $\alpha/2$ of half periods of  $\sin\left(\frac{\pi\alpha(d-1)}{4}\right)$ implying that $\Delta_{\alpha} (d)$ will change sign $(\alpha/2-1)$ times  for d within each such interval.  The Casimir amplitude $\Delta_{\alpha}(d)$ is in Fig.~1 plotted as function of $d$ for $\alpha\in\{2,4,6\}$. This illustrates the oscillatory behavior accompanied  by increasing amplitude for growing $d$ at $d$ sufficiently large.   \\
 
 3) For $\alpha$ not being an even integer there will always be both zeroes and singularities of the Casimir amplitude $\Delta_\alpha (d)$ indicating that (at the singular points in the $d$-space) the surface free energy is not well defined. For example, for $\alpha=3$ the singularities of $\Delta_{3} (d)$ occur at $d\in\{3,7,11,\dots\}$.   \\
 \begin{figure}
 \begin{center}
 \includegraphics[width=14cm]{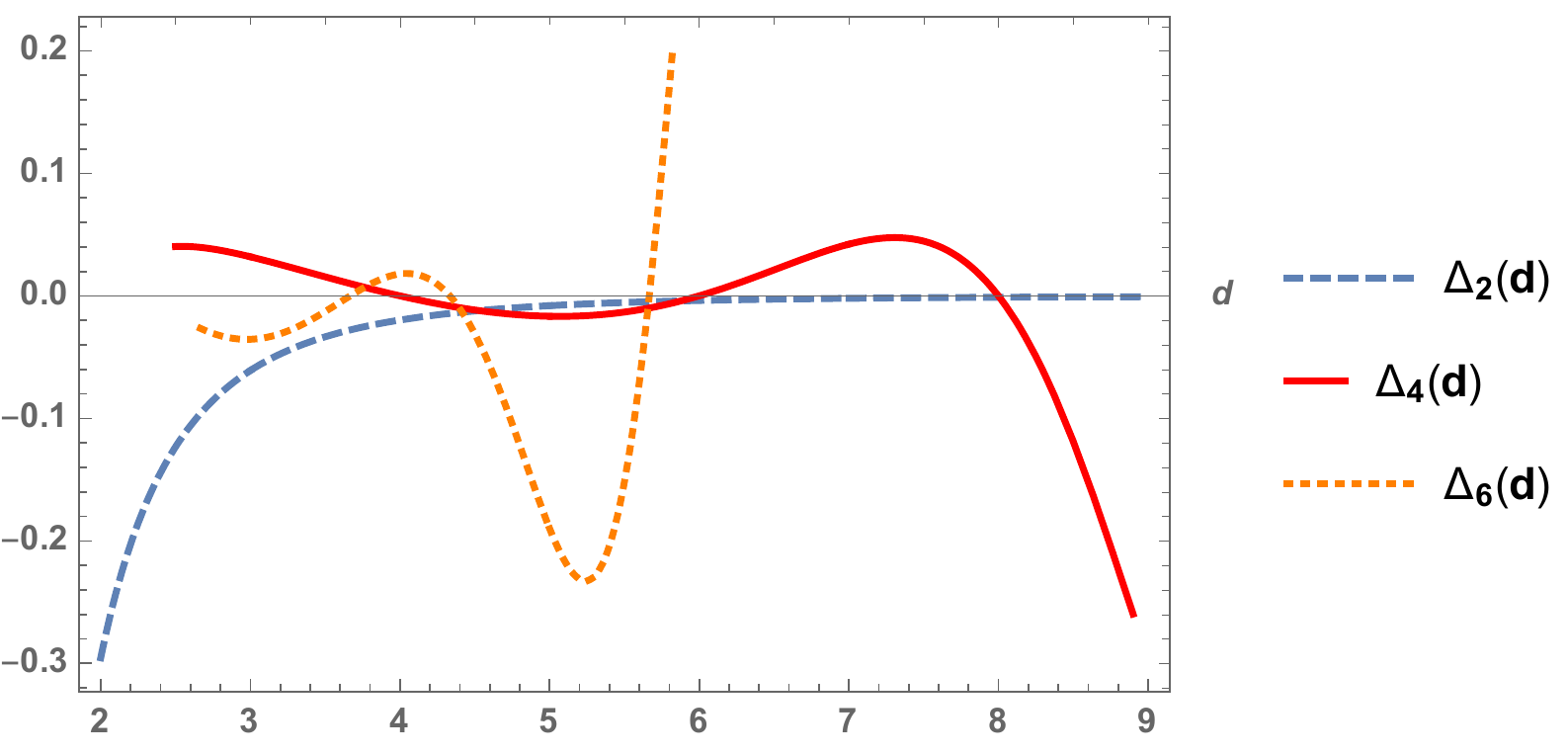} 
 \end{center}
\caption{The Casimir amplitude $\Delta_{\alpha}(d)$ plotted for a sequence of even values of $\alpha$. The function $ \Delta_{2}(d)$ remains negative and exhibits a negative maximum at $d\approx 21$ (not visible in this plot scale). For $d\to \infty$ the function $ \Delta_{2}(d)$ diverges towards negative infinity.  The corresponding force is always attractive in agreement with the general expectations. For $\alpha=4$ and $\alpha=6$ we observe an oscillatory behavior indicating alternations of the Casimir force sign upon varying dimensionality. The same holds true for higher even values of $\alpha$. For $\alpha\in\{4,6,8,\dots\}$ the magnitude of the oscillations violently diverges for large  $d$. The presented plot corresponds to a choice of the non-universal microscopic parameters ($t_0,t$) such that $\frac{\lambda_2^{\alpha/2}}{\lambda_1}=1$. Note that certain quantitative features of the plotted profiles, such as the position of the maximum of $ \Delta_{2}(d)$ depend on this choice.      } 
\end{figure} 
We emphasize that from the point of view of Eq.~(\ref{Cas_amp_fin}) the standard, physical case of $\alpha=2$ is an extremely special one. In this case all the singularities and zeroes in the expression for $\Delta_\alpha (d)$ that were mentioned above, conspire to yield a smooth curve with no zeroes.  

 We now adopt a complementary point of view on Eq.~(\ref{Cas_amp_fin}). We fix $d$ and follow the evolution of the Casimir amplitude when varying the parameter $\alpha$. It is clear that at fixed 
 $d\notin \{3, 5, 7,\dots\}$ the factor $\Gamma\left(\frac{1-d}{2}\right)$ in Eq.~(\ref{Cas_amp_fin}) remains finite and the function $\Delta_\alpha (d)$ evolves smoothly and oscillates upon varying $\alpha$. However, for odd values of $d$ there is no other way for compensating the singularity of  $\Gamma\left(\frac{1-d}{2}\right)$ than taking very special (natural even) values of $\alpha$. In particular the limit 
 $\lim\limits_{\alpha\to 2n}\,\lim\limits_{d\to 2m+1} \Delta_\alpha (d)$ for $n,m\in \{1,2,3,\dots\}$ does not exist. The limiting expression makes sense only if $\alpha$ is sent to  $2n$ before taking the limit $d\to 2m+1$. The dependence of $\Delta_{\alpha}(d)$ on $\alpha$ for a sequence of values of $d$ approaching $d=3$ is exhibited in Fig.~2. It illustrates the divergence of $\Delta_{\alpha}(d)$ for $d\to 3^+$ for all $\alpha \notin \{2,4,6,\dots \}$. 
 \begin{figure}
\begin{center}
 \includegraphics[width=14cm]{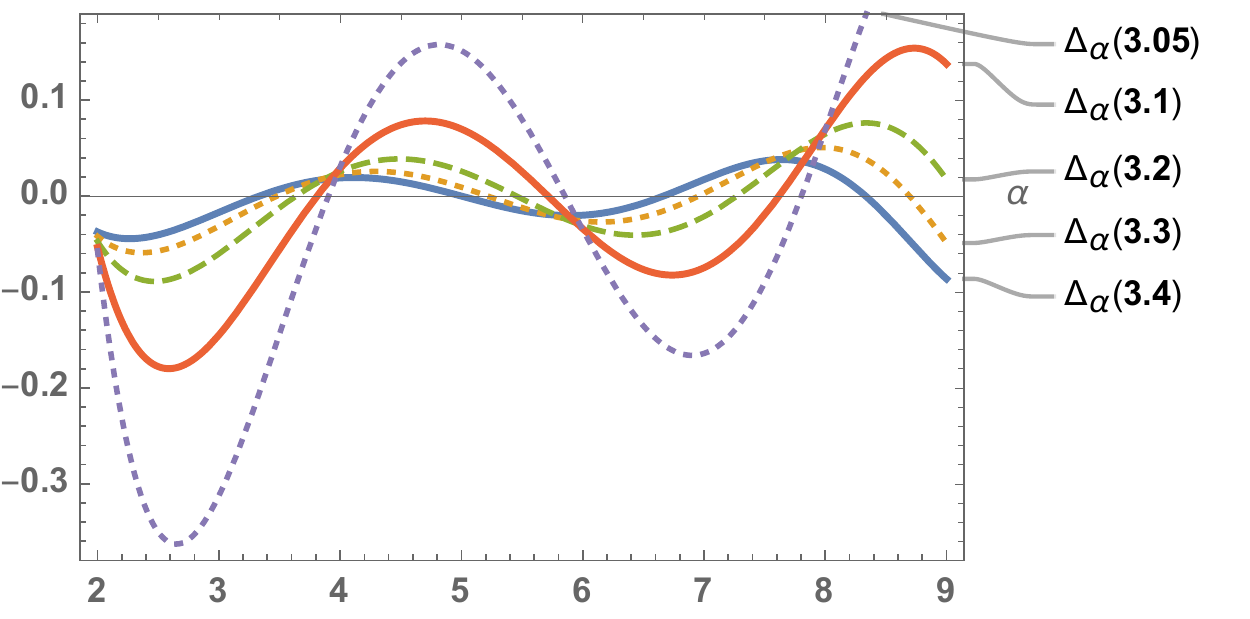}
\end{center} 
\caption{The Casimir amplitude $\Delta_{\alpha}(d)$ plotted as a function of $\alpha$ for a sequence of dimensionalities $d$ approaching $d=3$ from above. The function  $\Delta_{\alpha}(d)$ diverges for $d\to 3^+$ for any fixed value of $\alpha$ except $\alpha \in\{2,4,6,\dots \}$. The presented plot corresponds to a choice of the non-universal microscopic parameters ($t_0,t$) such that $\frac{\lambda_2^{\alpha/2}}{\lambda_1}=1$. }
\end{figure}
We finally illustrate the smoothening of the singularity of $\Delta_\alpha (d)$ for $\alpha$ approaching an even natural value. In Fig.~3 we plot a set of projections of $\Delta_\alpha (d)$ for $\alpha$ approaching 2.   
\begin{figure}
\begin{center}
 \includegraphics[width=14cm]{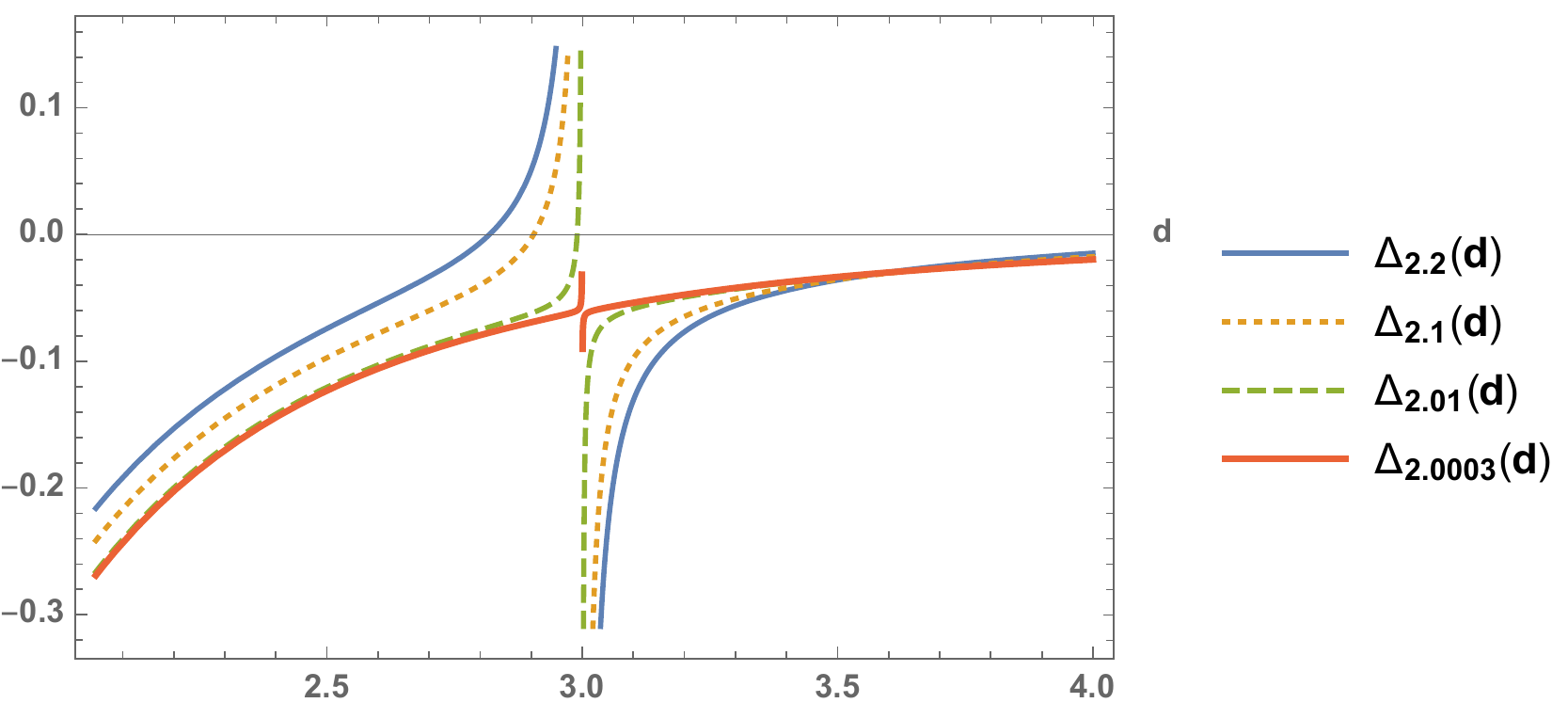}
\end{center} 
\caption{The Casimir amplitude $\Delta_{\alpha}(d)$ plotted as a function of $d$ for a sequence of values of $\alpha$ approaching $\alpha=2$ from above. The function  $\Delta_{\alpha}(d)$ diverges at $d=3$ for all the exhibited values of $\alpha$. An approach towards the limit $\alpha\to 2^+$, where the singularity is absent is illustrated. Upon reducing $\alpha$ towards two, the vicinity of the singularity becomes progressively squashed, and finally vanishes for $d\to 2$.  The presented plot corresponds to a choice of the non-universal microscopic parameters ($t_0,t$) such that $\frac{\lambda_2^{\alpha/2}}{\lambda_1}=1$.}
\end{figure}

It is also of interest to trace back the origin of this complex behavior depending on $\alpha$, which is encoded in the properties the function $\Psi_\alpha (z)$ defined by  Eq.~(\ref{Cas_en}), which is the Fourier transform of $ e^{-|x|^\alpha}$. For $\alpha=2$ the function $\Psi_\alpha(z)$ is gaussian, while for $\alpha\in \{4,6,8,\dots\}$ it may be related to generalized hypergeometric functions and exhibits exponentially damped oscillations at large arguments \cite{Boyd_2014}. However, as we verified numerically, for $\alpha\notin \{2, 4,6,8,\dots\}$ the function features a power-law tail, which has strong influence on the properties of  the integral in Eq.~(\ref{I_int_1}). 

It is also worth mentioning the behavior of the Casimir force in the complementary anisotropic case in which the dispersion in the direction perpendicular 
to the walls remains quadratic while it takes the form $|\kgr|^\alpha$ in one of the directions parallel to the walls. A straightforward calculation shows that in  this case the coefficient of surface tension takes the form 
\beq
\label{st10}
\overline{\sigma}^{N}(T,\mu) =  - \frac{k_{B}T}{4 \lambda_{1}^{d-2} \lambda_{2}}\,\Gamma\left(1+\frac{1}{\alpha}\right) \,g_{\frac{\alpha d+2}{2\alpha}}\left(e^{\beta\mu}\right) 
\eeq
and is negative while the analogue of Eq.~(\ref{Cas_en}) takes the following form
\beq 
\label{Cas_en_mod}
\frac{\delta\omega_s(T,0,D)}{k_B T} = \frac{\overline{\Delta}_\alpha (d)}{D^{\frac{\alpha(d-2)+2}{\alpha}}}\;, 
\eeq
where 
\beq 
\label{Cas_amp_mod}
\overline{\Delta}_\alpha (d) = - \frac{2^{-\frac{\alpha(d-2)+2}{\alpha}}}{\pi^{\frac{\alpha(d-1)+2}{2\alpha}}}\, \frac{\lambda_{1}^{\frac{2}{\alpha}}}{\lambda_{2}} 
\,\Gamma\left(1+\frac{1}{\alpha}\right)\,\Gamma\left(\frac{\alpha(d-1)+2}{2\alpha}\right)\,\zeta\left(\frac{\alpha(d-1)+2}{\alpha}\right)\,.
\eeq
It follows that in this case the Casimir force remains always attractive and the amplitude $\overline{\Delta}_\alpha (d)$ is a smooth function 
of  the parameters $\alpha$ and $d$. Again, in the special case $\alpha=2$ and $d=3$ one recovers the previous results for the coefficient of surface 
tension $\overline{\sigma}^N(T,\mu) = \sigma^N(T,\mu)$, see Eq.~(\ref{st1}), and the Casimir amplitude $\overline{\Delta}_2 (3) = \Delta_{2}(3) = -\zeta(3)/8\pi$.



\section{Summary}
In this paper we have analyzed the thermodynamic Casimir forces in an ideal Bose gas hosting the condensate and characterized by an anisotropic dispersion. The principal physical context for the study derives from ultracold Bose systems on optical lattices, where the anisotropic dispersion relations may be flexibly engineered. Our motivation is related to the recent studies of Refs.~\cite{Burgsmuller_2010, Lebek_2020, Lebek_2021}, reporting a far-going deviation of basic properties of the Casimir forces as compared to the usual, isotropic situations, in particular variation of the Casimir force sign depending on the dimensionality. The simplicity of the employed model allowed us to derive a closed analytical expression for the Casimir amplitude $\Delta_\alpha (d)$ as a function of the dimensionality $d$ and the exponent $\alpha$ governing the asymptotic decay of the dispersion for momentum ${\bf p}$ oriented perpendicular to the confining walls. The physically admissible values of $\alpha$ for the case of hypercubic lattices are restricted to natural even numbers ($\alpha\in \{2,4,6,\dots\}$). It is however interesting to consider $\alpha$ as a continuous variable akin to what is a common practice for the spatial dimensionality $d$. 

Our study confirmed the expected modification of the power-law governing the decay of the Casimir energy upon increasing the distance $D$. This modification is due to the anisotropy and the emergence of a dimensionful scale factor in the expression for the Casimir amplitude, which precludes its universality. It also provided an expression for the decay exponent for general values of $d$ and $\alpha$. Our formula for $\Delta_\alpha (d)$ fully agrees with earlier results, and in fact coincides with that reported for the imperfect Bose gas (at least for the cases of $\alpha\in\{2,4\}$, which were addressed for that system). In particular, it points at the variation of the sign of $\Delta_\alpha (d)$ upon changing $d$ at fixed $\alpha$. The obtained result for $\Delta_\alpha (d)$ (viewed as a function of $d$ and $\alpha$) hosts both zeroes and singularities. Quite surprisingly, recovering the standard results for $\alpha=2$ involves subtle cancellations of zeros and singularities and turns out to represent the unique case where one obtains a nonsingular function of $d$ characterized by a fixed sign. The situation for $\alpha\in\{4,6,8,\dots\}$ turns out to be different in that the apparent singularities of $\Delta_\alpha (d)$ cancel, but some zeroes remain, which gives rise to the oscillatory behavior of $\Delta_\alpha (d)$ as a function of $d$. For  $\alpha\notin\{2,4,6,8,\dots\}$ the expression for $\Delta_\alpha (d)$ always features singularities for some values of $d$. Our result implies that no pair of physically relevant situations (e.g. $\alpha=2$ and $\alpha=4$) can be continuously connected by varying $\alpha$ at fixed $d$. 

An interesting open question is to check the robustness of the presented picture with respect to interactions which are treated beyond the mean-field approach. Our predictions are presumably not easy to test experimentally at the present point, considering the required amount of tuning. We believe however that they are open to verification via numerical simulations.

\ack
PJ acknowledges support from the Polish National Science Center via grant 2014/15/B/ST3/02212.

\section*{References}

\bibliographystyle{iopart-num}
\bibliography{./refs}{} 

\providecommand{\newblock}{}


\end{document}